\documentclass[aoas,nameyear,dvips]{arximspdf}

\doi{10.1214/10-AOAS431}
\volume{4}
\issue{4}
\pubyear{2010}
\firstpage{1652}
\lastpage{1655}

\begin{document}
\begin{frontmatter}

\title{Selected recollections of my relationship with~Leo~Breiman}
\runtitle{My relationship with Leo Breiman}

\begin{aug}
\author[A]{\fnms{Charles J.} \snm{Stone}\ead[label=e1]{stone@stat.berkeley.edu}\corref{}}
\runauthor{C. J. Stone}
\affiliation{University of California, Berkeley}
\address[A]{Department of Statistics\\
 University of California, Berkeley\\
 Berkeley, California 94720\\
 USA\\
\printead{e1}} 
\end{aug}

\received{\smonth{10} \syear{2010}}



\end{frontmatter}

During the period 1962--1964, I had a tenure track Assistant Professorship
in Mathematics at Cornell University in Ithaca, New York, where I did
research in probability theory, especially on linear diffusion processes.
Being somewhat lonely there and not liking the cold winter weather, I
decided around the beginning of 1964 to try to get a job in the
Mathematics Department at UCLA, in the city in which I was born and raised.
At that time, Leo Breiman was an Associate Professor in that department.
Presumably, he liked my research on linear diffusion processes and other
research as well, since the department offered me a tenure track
Assistant Professorship, which I happily accepted. During the Summer
of 1965, I worked on various projects with Sidney Port, then at RAND
Corporation, especially on random walks and related material.
I was promoted to Associate Professor, effective in Fall, 1966,
presumably thanks in part to Leo. Early in 1966, I~was surprised
to be asked by Leo to participate in a department meeting called to
discuss the possible hiring of Sidney. The conclusion was that
Sidney was hired as Associate Professor in the department, as of
Fall, 1966. Leo communicated to me his view that he thought that
Sidney and I worked well together, which is why he had urged the
department to hire Sidney. Anyhow, Sidney and I had a very fruitful
and enjoyable collaboration in probability and, to a much lesser
extent, in theoretical statistics, for a number of years thereafter.

In 1967, Leo decided to leave academia in order to become a
full-time consultant. The purported reason, as I heard it, was that
he wanted to devote his attention to studying how children
tackle math problems. I then had virtually no contact with him
for a number of years. I did hear that he ran for and got elected
to the Santa Monica Board of Education. He was then elected President
of the Board. This constrained his available time for consulting
at Technology Service Corporation, where he was now their full time
consultant. So I got a call from TSC asking if I would like to
consult for them, which I did.

At TSC, Breiman had been working with its employees (mainly John Gins,
a~programmer and statistician) and with Jerry Friedman on tree-structured
classification and regression. I joined that effort. Previous efforts,
such as AID, had used the hypothesis testing framework to determine
when to stop growing a tree. I wasn't exactly enamored with this approach.
Breiman and I developed an alternative approach based on
tree growing, followed by tree pruning, by analogy with stepwise
addition followed by stepwise deletion in multiple regression.
Breiman and I wrote up our ideas, including the optimal pruning
algorithm, in a technical report that was never submitted
for publication. However, these ideas played a fundamental role in
the monograph Classification Trees (CART) by Breiman,
Friedman, Olshen and Stone, which was published in \citeyear{Betal1984}.
(CART is now a trademark of California Statistical Software, Inc.
and corresponding software is currently available from Salford Systems.)
The optimal pruning algorithm is also known as the BFOS algorithm.

In the preface to the CART book it is stated that its conception
occurred in 1980 and that ``While the pregnancy has been rather
prolonged, we hope that the baby appears acceptably healthy
to the members of our statistical community.'' Here are some
further details about the writing of the book. Friedman was largely
responsible for the computer/software effort. As far as the writing
was concerned, it was agreed that Leo would write a first draft of the book
and then Olshen and I would have a take at it. However, when our revision
didn't emerge after some months,  Leo sulked and stated that he wouldn't
interact with us about our rewriting. Whenever we finished it,
he would take a look and decide what to do at that time. When
he did see it, he didn't like it at all. We were at a standstill
for a considerable amount of time. Finally, it was decided to
create the book more or less by using his version followed by
our version. Thus Chapters 1--5, 7 and 8 of Breiman et al. (\citeyear{Betal1984})
were basically his version, while Chapters 6 and 9--12 were basically the OS
version.

In mid-1989 Padraic Neville, a CART programmer and statistician,
and I took on a consulting project with a relatively small oil company,
the purpose of which was to use a possibly modified version of CART
to develop a system for generating signals for trading
oil and currency futures based on a variety of time series data
supplied by the customer. Early on in the project, we realized
how sensitive our signals were to slight pertubations in the
data, such as small changes to the length of the time series,
the individual variables, or the details of cross-validation.
In short, the signals were highly discontinuous or unstable functions of the
data. This reminded me of a couple of a very brief conversation
that I had previously had with Charles Stein. I asked him his
opinion of stepwise or all subsets regression. His answer was
that he didn't like such procedures since any admissible procedure
should be an analytic function of the data. He also said that
he would be working on better alternatives in the not-too-distant
future. (Later I checked back with him and he hadn't gotten
around to that yet.) Anyhow, if a procedure should be an analytic
function of the data, then, at least, it should be a continuous
function of the data and, certainly, not a highly discontinuous
or unstable function. In order to reduce the instability of CART
in the intended application, we tried making various perturbations
of the data and methodology---especially, changing the length of the time series
or changing the cross-validation procedure, generating the resulting regression
or probability trees, and averaging the results. Part of my heuristic
reasoning was that differences in predictors produced by regression
trees based on slight perturbations in the data or methodology
should mainly be due to noise in the predictors rather than differing amounts
of bias, so averaging the resulting predictors should help.
This approach indeed seemed to improve the stability of the resulting signals,
but in early 1991 the customer canceled the project, claiming that its results
were inferior to other approaches it was using. Among the contributing
factors to our lack of success at the time might have been Iraq's
invasion of Kuwait in mid-1990, which had a very disruptive effect
on the time-series data we were using at the time. After the project ended,
Neville and I discontinued our work on combining CART trees, one
main reason being that he had no interest in (unpaid) academic
research. However, in the Summer of 1991 I had some research funds to
support a Ph.D. student, but no student who needed such support,
while Leo had a Ph.D. student, Samarajit Bose, who had no financial support
that summer. So I hired Bose on my grant to use simulated data
to study the effectiveness of averaging predictors in the contexts
of regression trees and stepwise regression. It was apparent that the averaging procedure produced significant benefits in both contexts, but more so
in the former context, presumably because CART is significantly more unstable
than stepwise regression. Somewhat later, I had a brief conversation
with Leo, in which I told him essentially this story. I do not
know what influence this conversation may have had on Leo's later research,
especially that on bagging predictors, the heuristics of instability
in model selection, and arcing classifiers.

While Leo and I were acting as TSC consultants in the late 1970s,
we, together with John Gins, got involved with extreme upper quantiles
in the context of air pollution data. Suppose, in particular, that one has
a series of data maxima of air pollution data, except that a small
proportion of the daily maxima are missing, and the goal is to estimate
the second highest daily maximum during the previous year (which figured into
federal regulations at the time). This goal is closely related to that of
estimating the extreme upper quantiles of the daily maxima. Anyhow,
the three of us wrote several technical reports on estimates
and confidence intervals for extreme upper quantiles. After
Leo and I moved to Berkeley, we decided to submit a paper on
confidence intervals for extreme upper quantiles. I was the one who
mainly was involved with writing and submitting the paper, but it
was based on the research that Leo was heavily involved with at TSC.
Our paper was tentatively accepted by \textit{JASA}. However, by the time I
got around to revising it as recommended, there was a change in the
editorial staff at \textit{JASA} and the revised version was rejected. I
then submitted a revised revised version to \textit{Technometrics}, but it too
was rejected. My diagnosis of the problem is that our paper was
being reviewed by specialists in extreme value theory who were
guarding their turf. Anyhow, time passed and I got involved in
other activities, including logspline density estimation. I had
a Ph.D. student, Charles Kooperberg, who I was supporting as an RA
on my NSF grant. He and I together worked on the problem of trying
to use logspline density estimation to yield competitive confidence
intervals for extreme upper quantiles, with the goal of incorporating
the resulting procedure into my paper with Leo. Ultimately, however,
the confidence interval procedure based on logspline density estimation
did not perform as well as another of our procedures, the
quadratic tail procedure (which was actually developed by Leo
in one of our TSC technical reports). Thus I wrote up, with
programming support by Kooperberg, another version of my paper with
Leo and submitted it to the \textit{Journal of Statistical Computation
and Simulation}. Fortunately, it was accepted without much difficulty
and published as Breiman, Stone and Kooperberg (\citeyear{BSK1990}). Kooperberg went on to
get his Ph.D. in 1991 with David Donoho as his main advisor but with
one of the chapters in his dissertation on log-spline density estimation.
Kooperberg and I collaborated on many papers after that. Unfortunately,
however, for one reason or another, Leo and I never collaborated
on any research projects after the publication of our paper
on extreme upper quantiles. Also, Leo was very much turned off
by the editorial policy of \textit{JASA}.

\printaddresses

\end{document}